# Nonlinear Electrophoresis of Dielectric and Metal Spheres in a Nematic Liquid Crystal

Oleg D. Lavrentovich, Israel Lazo, Oleg P. Pishnyak

*Liquid Crystal Institute and Chemical Physics Interdisciplinary Program,*
*Kent State University, Kent OH, 44242.*

Electrophoresis is a motion of charged dispersed particles relative to a fluid in a uniform electric field [1]. The effect is widely used to separate macromolecules, to assemble colloidal structures, to transport particles in nano- and micro-fluidic devices and displays [2,3,4]. Typically, the fluid is isotropic (for example, water) and the electrophoretic velocity is linearly proportional to the electric field. In linear electrophoresis, only a direct current (DC) field can drive the particles. An alternate current (AC) field is more desirable because it allows one to overcome problems such as electrolysis and absence of steady flows [5,6]. Here we show that when the electrophoresis is performed in a nematic fluid, the effect becomes strongly non-linear with a velocity component that is quadratic in the applied voltage and has a direction that generally differs from the direction of linear velocity. The new phenomenon is caused by distortions of the LC orientation around the particle that break the fore-aft (or left-right) symmetry. The effect allows one to transport both charged and neutral particles, even when the particles themselves are perfectly symmetric (spherical), thus enabling new approaches in display technologies, colloidal assembly, separation, microfluidic and micromotor applications.

The electric charge of a particle dispersed in a fluid is screened by a diffuse cloud of mobile ions, called counterions, with charges of the sign opposite to that of the particles.

When an electric field is applied, the counterions and the particle move in opposite directions. For a small particle, the electrophoretic velocity **v** is determined by the electrostatic "pulling" force proportional to the applied electric field **E** and by the viscous drag force, as expressed by the Smoluchowski's formula [1]:

$$\mathbf{v} = \mu \mathbf{E}, \qquad (1)$$

where $\mu = \varepsilon_m \zeta / \eta$ is the electrophoretic mobility of the particle, $\zeta$ is the "zeta-potential" characterizing the charge of particle and its spatial distribution, $\varepsilon_m$ is the dielectric permittivity of the medium, $\eta$ is the medium's viscosity. According to Eq. (1), an AC field with a zero time average produces no net propulsion: A change in field polarity changes the sign of **v** and the period-averaged displacement is zero. This is why numerous applications of electrophoresis, ranging from translocation and separation of macromolecules [2] to controlled assembly of colloidal particles [3], microfluidics and displays [4], rely on a DC driving. The latter represents a certain drawback as the DC field might lead to undesirable electrochemical reactions, and explains a strong interest in mechanisms with nonlinear relationship between **v** and **E** [5,6]. Most studies consider isotropic fluids as an electrophoretic medium, in which case the nonlinear behavior can be observed either for high voltages or for particles with special properties, such as a patterned surface [6]. In the first case, the non-linear correction is cubic, $v = \mu E + \mu_3 E^3$, where $\mu_3$ is a non-linear mobility; a non-zero $\mu_3$ was found also in nematic liquid crystals [7].

We demonstrate a principally new type of electrophoresis in an orientationally ordered fluid, a nematic liquid crystal (LC), in which the dependence $\mathbf{v}(\mathbf{E})$ has a component *quadratic* in $E$. The dependence $v \sim E^2$ allows one to move particles even by a symmetric AC field with a modest amplitude, as the change in field polarity does not change $v$. The new phenomenon is caused by asymmetric distortions of LC orientation around the particle. The electrophoretic velocities linear and quadratic in $E$ generally have different directions, resulting in a high degree of freedom in moving the particle in space. We demonstrate electrophoretic motion parallel (antiparallel) to $\mathbf{E}$, perpendicular to $\mathbf{E}$, as well as motion along curvilinear tracks set by spatially varying orientation of the liquid crystal.

We used a nematic E7 (*EM Industries*) that melts into an isotropic phase at $t_{NI} = 58^0 C$. Molecular orientation in LC is described by the director $\hat{\mathbf{n}}$; since the medium is non-polar, $\hat{\mathbf{n}} = -\hat{\mathbf{n}}$. The LC layer of thickness $h = (50\text{-}80)\,\mu\text{m}$ between two glass plates is aligned uniformly along the *x*-axis, $\hat{\mathbf{n}}_0 = (1,0,0)$ by buffed polyimide layers. The field $\mathbf{E} = (E,0,0)$ is parallel to the *x*-axis. The dielectric anisotropy of E7 $\Delta\varepsilon = \varepsilon_{\|} - \varepsilon_{\perp} = 13.8$ is positive ($\varepsilon_{\|}$ and $\varepsilon_{\perp}$ are the dielectric permittivities for $\mathbf{E}$ parallel and perpendicular to $\hat{\mathbf{n}}$, respectively), so that $\mathbf{E}$ does not influence $\hat{\mathbf{n}}$ away from the particles. Two aluminum stripes separated by a gap *L*=5-12 mm served as the electrodes.

We used dielectric spheres of diameter $2a = (5-50)\mu\text{m}$ made of silica (*Bangs Laboratories*), borosilicate and soda lime glass (*Duke Scientific*), as well as gold spheres with $2a = (5.5-9)\mu\text{m}$ (*Alfa Aesar*). To ensure perpendicular orientation of $\hat{\mathbf{n}}$ at the particle's surface, the gold spheres were etched with an acid, while the dielectric spheres were functionalized with octadecyltrichlorosilane (OTS) or *N,N*-didecyl-*N*-methyl-(3-trimethoxysilylpropyl) ammonium chloride (DDMAC), both purchased from *Sigma-Aldrich*.

Ionic impurities make the voltage profile across the LC cell time-dependent. To screen the field, the ions move and build electric double layers near the electrodes, within a characteristic time [8] $\tau_e = \lambda_D L / 2D \approx (1-3)\min$, where $\lambda_D = (0.1-1)\mu m$ is the Debye screening length and $D = 10^{-10} - 10^{-11} m^2/s$ is the diffusion coefficient. We measure **v** within a few minutes of voltage application, in the regime of stationary motion. Then the voltage polarity is reversed and **v** is measured again.

In the isotropic melt, the dielectric spheres show a classic linear electrophoresis, Fig. 1a, similar to the observations of Tatarkova et al [9] made for both the isotropic melt and for an unaligned LC. The gold spheres do not move, $\mu_{Au}^I = 0$.

Once the material is cooled down into the nematic phase, each particle generates a radial director configuration in its immediate vicinity. To match the overall uniform alignment of the LC, each sphere acquires a satellite topological defect: either a point defect, the so-called hyperbolic hedgehog [10], Fig. 1d, or an equatorial disclination ring [11]. The pair hedgehog-sphere represents an elastic dipole [10] $\mathbf{p} = (p_x, 0, 0)$ that is elastically repelled from the bounding plates of the cell. The particles levitate in the bulk [12], thus resisting sedimentation which hinders electrophoresis in isotropic fluids [13]. If there is no field, the levitating spheres experience Brownian motion with zero net displacement.

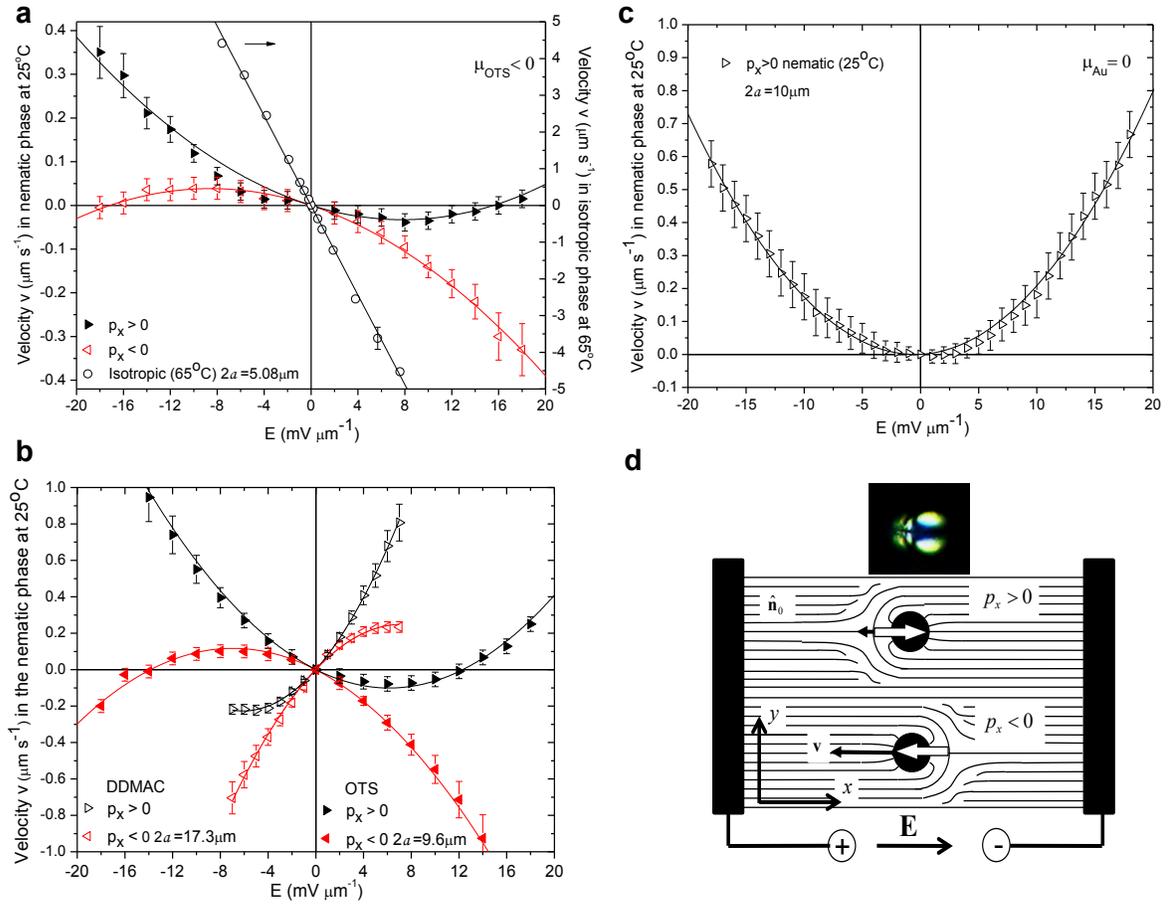

**Figure 1.** Nonlinear electrophoresis in DC field. Electrophoretic velocity vs applied field for: (a) OTS-coated silica spheres, $2a = 5.08\,\mu m$ in the isotropic phase ($65^0 C$) and in the nematic phase ($25^0 C$) of E7 with two polarities of the elastic dipole; (b) negatively charged OTS-coated borosilicate spheres, $2a = 9.6\,\mu m$, and positively charged DDMAC-coated borosilicate spheres, $2a = 17.3\,\mu m$, in the nematic phase; (c) neutral gold spheres in the nematic phase; (d) Scheme of experiment with $\mathbf{E}$ parallel to $\hat{\mathbf{n}}_0$; the hyperbolic hedgehog is either on the left-hand side of the sphere ($p_x > 0$) or the right ($p_x < 0$). The inset shows the polarizing microscope image of a glass sphere with $2a = 50\,\mu m$ and $p_x > 0$. Here and in other figures the error bars represent s.d. taken for over 100 particles.

Once the DC field is applied, the spheres with dipolar configurations demonstrate a strongly nonlinear electrophoresis, Fig. 1a,b,c with the velocity-field dependence

$$v = \mu E + \beta E^2 \tag{2}$$

For neutral gold particles, $\mu_{Au}^N = 0$ and the dependence $v(E)$ is parabolic, with $\beta_{Au}^N = 2 \times 10^{-3} \, \mu m^3 / mV^2 \cdot s$, Fig. 1c. For dielectric particles, both $\mu$ and $\beta$ are nonzero, e.g., $\mu_{OTS}^N = -0.03 \, \mu m^2 / mV \cdot s$, $\beta_{OTS}^N = 2.55 \times 10^{-3} \, \mu m^3 / mV^2 \cdot s$ for $2a = 9.6 \, \mu m$, and $\mu_{DDMAC}^N = 0.07 \, \mu m^2 / mV \cdot s$, $\beta_{DDMAC}^N = 5.5 \times 10^{-3} \, \mu m^3 / mV^2 \cdot s$ for $2a = 17.3 \, \mu m$. The nonzero quadratic term in Eq. (2) means that the electrophoresis in the LC can be naturally induced by an AC field, as confirmed experimentally, Fig. 2.

Equation (2) is written for the case when the vectors $\mathbf{v}$, $\mathbf{p}$, and $\mathbf{E}$ are all parallel (or antiparallel) to the x-axis, Fig. 1d. Generally, the dependence $\mathbf{v}(\mathbf{E})$ involves tensorial coefficients that depend on $\mathbf{p}$, so that the components of the velocity $v_i$ and the field $E_j$ ($i, j = x, y, z$) are related as follows:

$$v_i = \mu_{ij} E_j + \beta_{ijk} E_j E_k \; ; \tag{3}$$

The tensor character of $\mathbf{v}(\mathbf{E})$ is manifested by the fact that $\beta$ (but not $\mu$) changes sign with $\mathbf{p}$, so that the velocity components originating in the linear ($\mu_{xx} E_x$) and quadratic ($\beta_{xxx} E_x^2$) parts of Eq. (3) can be not only parallel but also antiparallel to each other, Fig. 1a,b, and 2a,b,d.

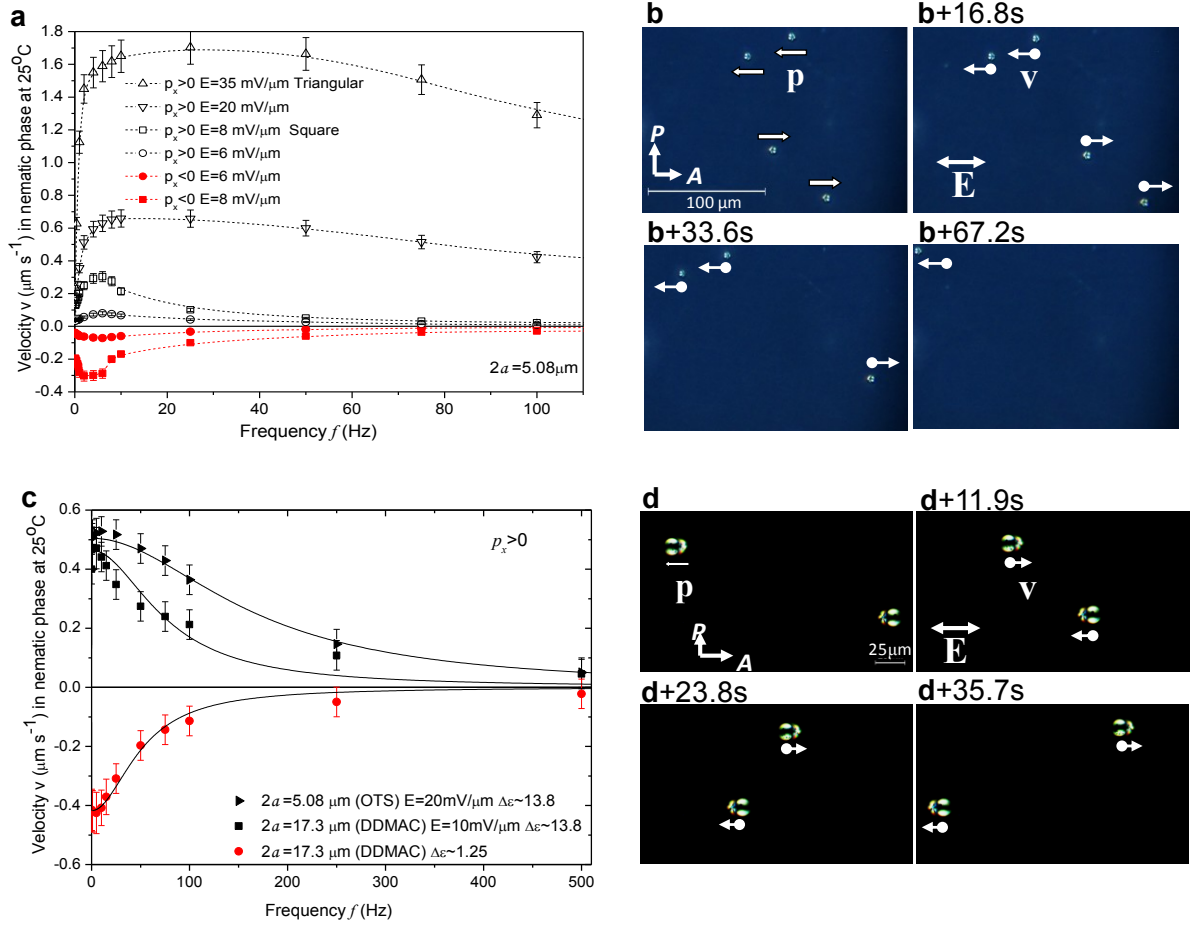

**Figure 2.** AC electrophoresis in the nematic LC: (a) OTS-coated silica particles, $2a = 5.08\,\mu m$, driven by triangular and square AC pulses. The lines are guides for eye; (b) Polarizing microscope textures of particles at time 0, 16.8; 33.6; 67.2 s after the triangular AC field $45\,mV/\mu m$, 100 Hz is applied; (c) The same as (a), but for the sinusoidal field; DDMAC-treated borosilicate particles, $2a = 17.3\,\mu m$, in a field $10\,mV/\mu m$ moving in E7 (squares), in a mixture with 18.7 wt% of E7 (circles) and OTS-treated silica particles with $2a = 5.08\,\mu m$ in a field $20\,mV/\mu m$ moving in E7 (triangles). The solid curves represent the fits with Eq. 4; (d) The same as (b), but for DDMAC-treated particles with $2a = 17.3\,\mu m$ in a mixture with 13.5 wt% of E7, and a field $30\,mV/\mu m$, 1 Hz.

Another illustration comes from the experiments with the LC MLC7026-000 (*Merck*) in which $\Delta\varepsilon = \varepsilon_\| - \varepsilon_\perp = -3.7$. We apply the field $\mathbf{E} = (0, 0, E_z)$ perpendicularly to $\mathbf{p} = (p_x, p_y, 0)$ (using transparent electrodes at the glass plates); since $\Delta\varepsilon < 0$, the field does not perturb the LC far away from the particles. By buffing the polyimide aligning layer in a circular fashion, we prepared a cell in which $\hat{\mathbf{n}}_0(x, y)$ formed a race-track configuration, Fig. 3. The spheres moved in the $(x, y)$ plane of the cell, perpendicular to $\mathbf{E}$, either counterclockwise or clockwise around the track, depending on the polarity of $\mathbf{p}$, Fig. 3.

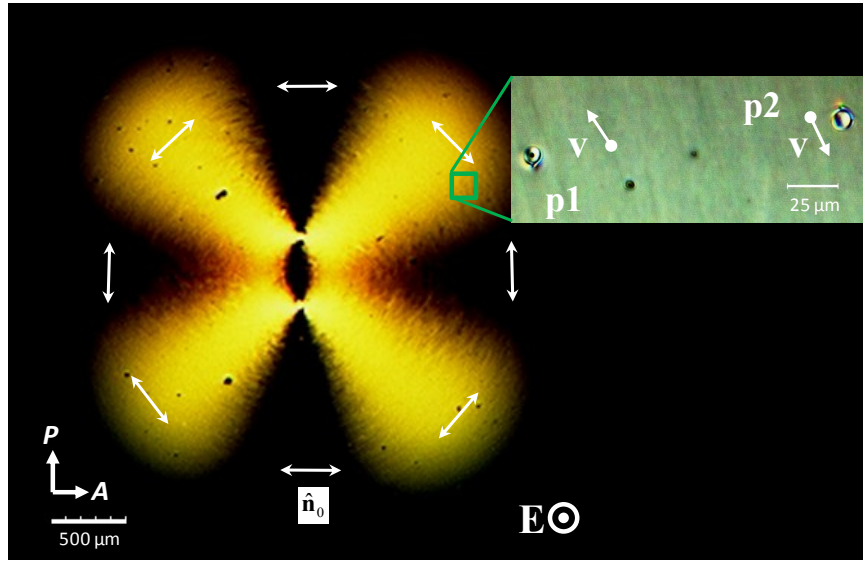

**Figure 3.** Electrophoretic motion of DDMAC-coated borosilicate spheres, $2a = 9.6\,\mu\text{m}$ in the plane perpendicular to the electric field, along the "race-track" trajectories set by a non-uniform director $\hat{\mathbf{n}}_0$.

In a separate experiment, we verified the role of dielectric anisotropy $\Delta\varepsilon$. In E7, $\Delta\varepsilon = 13.8$ is large, causing a dielectric torque $\propto \Delta\varepsilon$ on the director near the spheres. To minimize this torque, we mixed E7 with MLC7026-000; $\Delta\varepsilon$ reduces to 1.25 at the concentration 18.7 wt% of E7, and practically vanishes, $\Delta\varepsilon = 0.03$, at 13.45 wt% (measured for $25^0 C$ and 1kHz). The electrophoresis in these two mixtures was similar to the case of E7, Fig. 2c, demonstrating that the dielectric reorientation of $\hat{\mathbf{n}}$ is not the prevailing driving mechanism.

The AC and nonlinear DC electrophoresis are observed for spherical particles when the director distortions around them are of a dipolar type, Fig. 1d. If the director distortions preserve the fore-aft symmetry, as is the case of the equatorial defect ring, Fig. 4a, these effects vanish. To produce the defect structure with an equatorial ring, we followed a procedure from Gu and Abbott [14], namely, we used shallow cells with the separation between the plates that was close to the diameter of particles. The AC field caused back-and-forth linear electrophoresis of the particles but no net propulsion.

The experiments suggest that the mechanisms of the AC and non-linear DC electrophoresis are rooted in the type of director distortions of the carrier LC medium. In an isotropic fluid, the broken symmetry of *the particles themselves* can lead to a non-linear DC and AC electrophoresis [15-17] with $v \propto E^2$, as reviewed by Murtsovkin [13] and by Todd and Bazant who proposed a term "induced charge electrophoresis" (ICEP) [15]. The ICEP was experimentally demonstrated for anisotropic quartz particles [16] and for Janus spherical particles in an isotropic fluid [17]. The important difference of the LC electrophoresis is that the motion is caused by the broken symmetry of the medium rather than of the particle itself, Fig. 4.

Consider an uncharged metallic (gold) sphere in a LC. Once the field is on, the mobile ions of LC move in two opposite directions. These ions cannot penetrate the particle and thus accumulate at its opposite sides. The field-induced ionic clouds attract the "image charges" from within the conducting sphere, producing a Debye screening layer. The field-induced zeta potential can be estimated [15] as $\zeta_{ind} \sim aE$. In the steady state, the tangential component of **E** drives the mobile ions and thus the fluid from the poles of the sphere towards its equator. The directionality of this ICEP flow does not change with the field reversal, Fig. 4. If the particle and the surrounding medium have a mirror-image symmetry, the slip velocity produces no electrophoretic propulsion, as illustrated in Fig. 4a for a sphere with an equatorial defect ring. The hedgehog configuration, Fig. 4b, breaks this symmetry. The flows on the opposite sides of the sphere are not symmetric anymore and give rise to an electrophoretic velocity ~ $v \sim (aE)E \sim aE^2$.

This qualitative picture suggests that in Eq.(2), $\beta = \delta \varepsilon_m a / \eta$ for the metallic and $\beta = \delta \varepsilon_d \lambda_D / \eta$ for the dielectric particles; here $\delta$ is the dimensionless factor characterizing the medium asymmetry; $\delta = 0$ in Fig. 4a and $\delta \neq 0$ in Fig. 4b. Using $2a = 10\,\mu m$, $\eta = 0.1\,Pa \cdot s$ [9], $\varepsilon_m = 10\varepsilon_0$, $\varepsilon_d = 5.8\varepsilon_0$, $\lambda_D = 1\,\mu m$ and $\delta = 1$, one finds $\beta = 5 \times 10^{-3}\,\mu m^3 / mV^2 \cdot s$ for gold particles and $\beta = 0.5 \times 10^{-3}\,\mu m^3 / mV^2 \cdot s$ for the dielectric ones, of the same order as the experimental values.

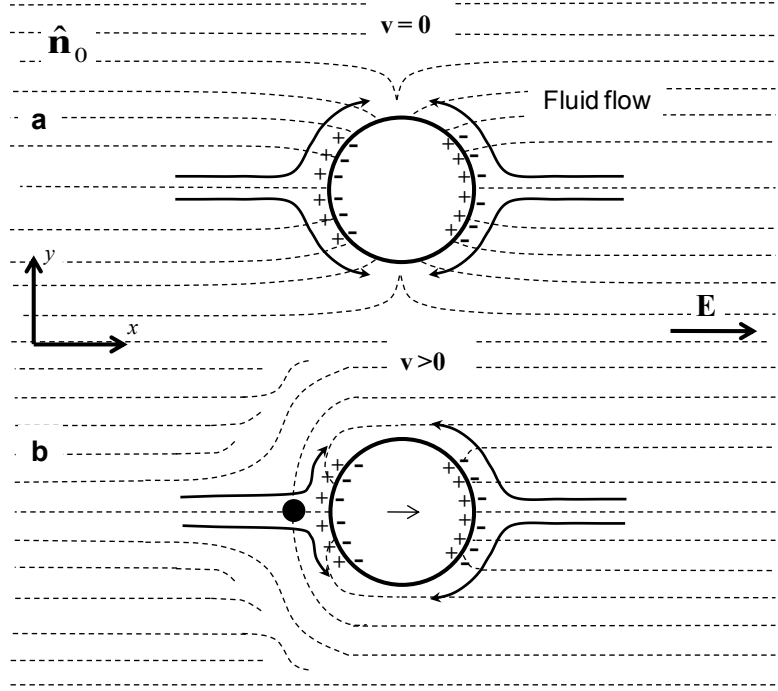

**Figure 4.** Spherical particles with normal anchoring in a nematic LC with (a) quadrupolar and (b) dipolar symmetry of director distortions. The Saturn ring (a) preserves the fore-aft symmetry and results in zero electrophoretic mobility; the hyperbolic hedgehog (b) breaks the fore-aft symmetry and is responsible for the nonlinear electrophoresis.

It is of interest to discuss how $v$ depends on the frequency of the sinusoidal AC field, Fig. 2c, following the considerations for isotropic fluids [15,18,19]. The ICEP velocity is controlled by two time scales; 1) a characteristic charging time $\tau_c = \lambda_D a / D$ (for a conductive sphere) or $\tau_c = \varepsilon_m \lambda_D^2 / \varepsilon_d D$ (for a dielectric sphere); 2) a characteristic electrode charging time $\tau_e = \lambda_D L / 2D$. In the order of magnitude, $\tau_c = 10^{-2}$ s and $\tau_e = 10^2$ s, with $\lambda_D = 1$ μm and $D = 10^{-11}$ m$^2$/s. For $\lambda_D \ll a \ll L$, the bulk AC field is controlled by $\tau_e$: $E_o(t) = \frac{V_o}{L} \cos(\omega t) \text{Re}\left[\frac{i\omega\tau_e}{1+i\omega\tau_e} e^{-i\omega t}\right]$. The time dependent polarization of the sphere is proportional to $\text{Re}\left(\frac{e^{i\omega t}}{1+\omega\tau_c}\right)$. The resulting frequency dependence

$$v(\omega) = v_o \frac{\omega^2 \tau_e^2}{(1+\omega^2\tau_c^2)(1+\omega^2\tau_e^2)}, \qquad (4)$$

fits the experimental data $v(f)$, where $f = \omega/2\pi$, very well, Fig. 2c. The velocity increases as $\omega^2$ when $\omega$ is low, but at the high $\omega$, $v$ decreases since the ions cannot follow the rapidly changing field. All three experimental dependencies in Fig. 2c were fitted with practically the same values of parameters in Eq. 4, namely $\tau_c$ in the range $(0.005-0.015)$s, $\tau_e$ in the range $(43-53)$s and $L=10$ mm.

To conclude, we demonstrate a highly nonlinear electrophoretic motion of metallic and dielectric spheres in a nematic LC. The particles can be driven either by a DC or AC field regardless of whether their zeta potential is finite (dielectric spheres) or zero (metallic spheres). The LC electrophoresis is much richer than its isotropic counterparts. In an isotropic fluid, the electrophoretic particle must be charged or be asymmetric. In LC electrophoresis, the particle can be of any shape, including a highly symmetric sphere and be of any charge (including zero). The components of velocity that originate in the linear and quadratic terms in Eq. (3) need not to be parallel to each other and the particles can be moved in any direction in 3D space. The described phenomenon offers new perspectives for practical applications where a highly flexible, precise and simple control of particle (or cargo) placement, delivery, mixing or sorting is needed. Examples include microfluidic devices and electrophoretic displays[20]. The practical potential of the LC-based electrophoresis is further expanded by the fact that the trajectories and velocities of particles can be controlled not only by the frequency dependent linear and quadratic mobilities in Eqs. (2,3), but also by the spatially varying director field $\hat{\mathbf{n}}_0(\mathbf{r})$.

**Acknowledgements** We are thankful to L. Tortora for help with surface functionalization of the particles and discussions and to the anonymous Referee for the constructive suggestions. The research was supported by NSF DMR.

**Author contributions.** Experimental strategy was designed by O.D.L. Initial DC and AC field experiments for dielectric spheres were performed by O.P.P. DC and AC field experiments for dielectric and metallic spheres, experiments with nonuniform director and analysis of data were performed by I.L. The paper was written by O.D.L.

**Author information** Correspondence and request for materials should be addressed to O.D.L. (olavrent@kent.edu)